\documentclass{aa}
\usepackage{natbib}
\bibpunct{(}{)}{;}{a}{}{,}
\usepackage{epsf}
\newcommand{\be}{\begin{eqnarray}}
\newcommand{\ee}{\end{eqnarray}}
\begin{document}

\title{Large Scale Structure in the SDSS DR1 Galaxy Survey}
\author{A. Doroshkevich\inst{1,2}\and
D.L. Tucker\inst{3}\and
S. Allam\inst{3,4}\and
M.J. Way\inst{1,5}}
\institute{Theoretical Astrophysics Center, Juliane Maries Vej 30,
DK-2100 Copenhagen \O, Denmark\and
Keldysh Institute of Applied Mathematics, Russian Academy of Sciences,
125047 Moscow,  Russia\and
Fermi National Accelerator Laboratory, MS 127, P.O. Box 500, 
Batavia, IL 60510 USA\and
National Research Institute for Astronomy \& Geophysics, Helwan
Observatory, Cairo, Egypt\and
NASA Ames Research Center, Space Sciences Division, 
MS 245-6, Moffett Field, CA 94035, USA
}
\date{Received ..., 2003/Accepted ..., 2003}

\abstract{
The Large Scale Structure in the galaxy distribution is
investigated using The First Data Release of the Sloan Digital 
Sky Survey. Using the Minimal Spanning Tree technique we have
extracted sets of filaments, of wall--like structures, of galaxy
groups, and of rich clusters from this unique sample. The physical
properties of these structures were then measured and compared with
the statistical expectations based on the Zel'dovich' theory.

The measured characteristics of galaxy walls were found to be
consistent with those for a spatially flat $\Lambda$CDM
cosmological model with $\Omega_m\approx$ 0.3 and $\Omega_\Lambda
\approx$ 0.7, and for Gaussian initial perturbations with a 
Harrison -- Zel'dovich power spectrum. Furthermore, we found that 
the mass functions of groups and of unrelaxed structure elements 
generally fit well with the expectations from Zel'dovich' theory. 
We also note that both groups and rich clusters tend to prefer the
environments of walls, which tend to be of higher density, rather 
than the environments of filaments, which tend to be of lower density.
}

\maketitle

\section{Introduction}

With the advent of the Durham/UKST Galaxy Redshift Survey
\citep[DURS,][]{RaSh:96} and the Las Campanas Redshift Survey
\citep[LCRS,][]{ShLa:96}, the galaxy distribution on scales up to
$\sim$300 $h^{-1}$~Mpc could be studied. Now these investigations can
be extended using the public data sets from The First Data Release of
the Sloan Digital Sky Survey \citep[SDSS DR1,][]{StLu:02,Ab:03},
which contains redshifts for $\approx$ 
100\,000 galaxies in four slices for distances $D\leq 600 h^{-1}$Mpc. 

The analysis of the spatial galaxy distribution in the DURS and the
LCRS has revealed that the Large Scale Structure (LSS) is composed of
walls and filaments, that galaxies are divided roughly equally into
each of these two populations (with few or no truly isolated
galaxies), and that richer walls are linked to the joint random
network of the cosmic web by systems of filaments \citep{DoFo:00,DoFo:01}.
Furthermore, these findings are consistent with results
obtained for simulations of dark matter (DM) distributions
\citep[see e.g.][]{CoHa:98,JeFr:98} and for mock galaxy catalogues
based upon DM simulations \citep{CoHa:98}.

The quantitative statistical description of the LSS is in itself an
important problem. Beyond that, though, the analysis of rich
catalogues can also provide estimates for certain cosmological
parameters and for the characteristics of the initial power spectrum
of perturbations. To do so, some theoretical models of structure
formation can be used.

The close connection between the LSS and Zel'dovich' pancakes has been
discussed by \cite{ToGr:78} and by \cite{Oo:83}.  Now this
connection is verified by the comparison of the statistical
characteristics of observed and simulated walls with theoretical
expectations \citep[][hereafter DD99 and DD02]{DeDo:99,DeDo:02}
based on the Zel'dovich theory of nonlinear gravitational
instability \citep{Zeld:70,ShZeld:89}. This
approach connects the characteristics of the LSS with the main
parameters of the underlying cosmological scenario and the initial
power spectrum, and it permits the estimation of some of these
parameters using the measured properties of walls. It was examined
with the simulated DM distribution \citep[DD99;][]{DeDoTu:00},
and was found that, for sufficiently representative samples of walls,
a precision of better than 20\% can be reached.

Effective methods of the statistical description of the LSS 
based on the Minimal Spanning Tree (MST) technique were developed 
by \cite{DeDoTu:00} and \cite{DoFo:00,DoFo:01}, 
who applied them to DM simulations and to the DURS and the LCRS. 
These methods introduced in \cite{BaBhSo:85} and 
\cite{Weygaert1991} generalize the popular "friends-of-friends" 
approach. In this paper we apply the same approach to the SDSS DR1, a
sample from which we can obtain more representative and more precise
measures of the properties of the LSS and the initial power spectrum
of perturbations. Alternative methods based on Minkowski functionals 
are proposed by, for example, \cite{ShGo:99}, or \cite{ShSa:02}
 
\begin{figure}
\centering
\epsfxsize=7cm
\hspace{0.2cm}
\epsfbox{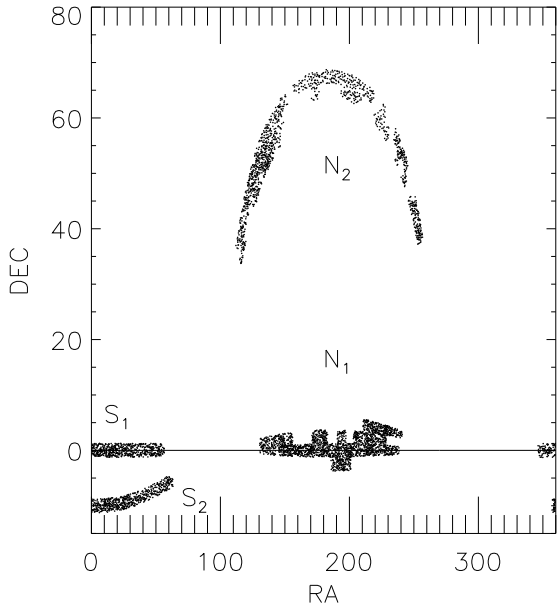}
\vspace{1.cm}
\caption{Four regions of the DR1 sample on the sky.
} 
\label{sky}
\end{figure}

With the MST technique, we can quantitatively describe the sample under
investigation and divide the sample into physically motivated subsamples 
of clusters with various threshold overdensities bounding them. Then we can
extract the LSS elements and characterize their morphology. In 
particular, this technique allows us to discriminate between filamentary 
and wall--like structure elements located presumably within low and high 
density regions and to estimate their parameters for the different 
threshold overdensities. The same technique allows us to extract sets 
of high density groups of galaxies and to measure some of their 
properties. 

The analysis of wall-like condensations is most informative.
Comparison of the observed characteristics of walls with theoretical 
expectations (DD99; DD02) demonstrates that the observed
galaxy distribution is consistent with Gaussianity initial
perturbations and that the walls are the recently formed, partly
relaxed Zel'dovich' pancakes. The mean basic characteristics of the
walls are consistent with those theoretically expected for the initial
power spectrum measured by the CMB observations summarized, for 
example, in \cite{Sper:03}.

In this paper we also analyse the mass functions of structure elements
selected for a variety of boundary threshold overdensities.  We show
that these functions are quite similar to the expectations of
Zel'dovich' theory, which generalizes the Press -- Shechter formalism
for any structure elements.  In addition, the theory indicates that the
interaction of large and small scale perturbations can be important
for the formation of the observed LSS mass functions. Our analysis
demonstrates that this interaction is actually seen in the influence
of environment on the characteristics of groups of galaxies.  
This problem was also discussed in \cite{Ee:03,Ein:03}. 

\begin{figure}
\centering
\epsfxsize=7cm
\hspace{0.5cm}
\epsfbox{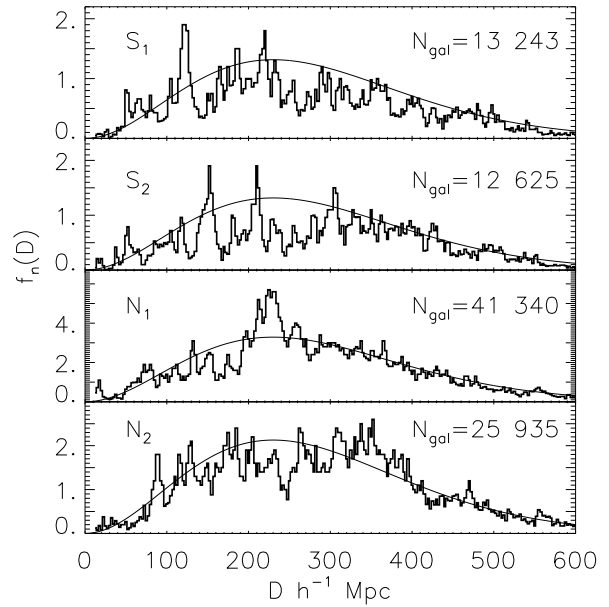}
\vspace{1.cm}
\caption{The radial galaxies distributions in the four samples 
of the SDSS DR1. The selection function (\ref{sel}) is plotted by 
solid lines.
} 
\label{hst}
\end{figure}

This paper is organized as follows: In Secs. 2 we describe the sample
of galaxies which we extracted from the SDSS DR1 and the method we
have employed to correct for radial selection effects.  In Sec. 3 we
establish the general characteristics of the LSS. More detailed
descriptions of the filamentary network and walls can be found in Secs. 4
and 5, respectively. In Secs. 6 and 7 we discuss the probable
selected clusters of galaxies and the mass function of structure
elements. We conclude with Sec.  8 where a summary and a short
discussion of main results are presented.

\section{The First Data Release of the Sloan Digital Sky Survey}

We use as our observational sample the SDSS DR1 \citep{Ab:03},
which is the first public release of data from the SDSS
\citep{Fu:96,Gunn:98,York:00}.  
The imaging data for the SDSS DR1 encompasses 2099~sq~deg of sky.
The DR1 also contains 186,240 follow up spectra, which are available over
1360~sq~deg of the imaging data area. Galaxies are situated within two 
north fields, $N_1\,\&\,N_2$, and two south fields, $S_1\,\&\,S_2$. 
These regions are plotted in Fig. \ref{sky}\,.

We obtained our SDSS DR1 sample via the SDSS DR1 Spectro query server
\footnote{ http://www.sdss.org/dr1}. This is a web interface to the SDSS 
Catalog Archive Server. We selected all objects identified as galaxies 
with a `redshift confidence minimal level' of 95\% and no maximal level. 
No other constraints in the selection were made at this level.

Our method for detecting LSS depends on having largely contiguous regions.
Hence we have removed some regions and artifacts from the DR1 sample.
The following are the RA and DEC areas we masked out from the original
DR1 query before our analysis:
\renewcommand{\labelitemi}{$\bullet$}
\begin{itemize}
\item 174$^{h}$$<$RA$<$179$^{h}$, -4.0$^{\degr}$$<$DEC$<$-1.22$^{\degr}$
\item 159$^{h}$$<$RA$<$163$^{h}$, 1.1$^{\degr}$$<$DEC$<$4.0$^{\degr}$
\item 10$^{h}$$<$RA$<$50$^{h}$, 10$^{\degr}$$<$DEC$<$20$^{\degr}$
\item 300$^{h}$$<$RA$<$355$^{h}$, -12$^{\degr}$$<$DEC$<$-4$^{\degr}$
\item 250$^{h}$$<$RA$<$270$^{h}$, 52$^{\degr}$$<$DEC$<$67$^{\degr}$
\end{itemize}

\begin{figure}
\centering
\epsfxsize=7cm
\hspace{0.5cm}
\epsfbox{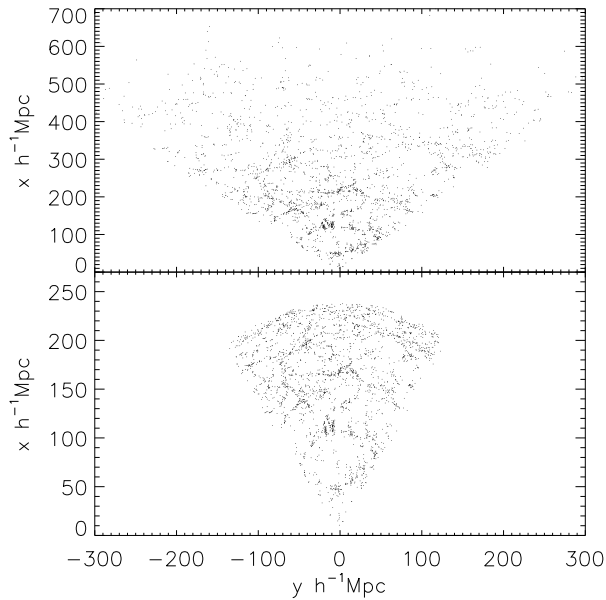}
\vspace{1.cm}
\caption{The wedge diagram of galaxy distributions in the samples 
$S_1$ is real (top panel) and modified (bottom panel) radial 
coordinates. 
} 
\label{wedge}
\end{figure}

\subsection{Correction for radial selection effects}

In Figure \ref{hst} we plot the radial distributions of galaxies in 
all four samples. In Figure \ref{wedge} we plot the wedge diagramm of 
observed galaxy distribution for the sample $S_1$. As is  clearly seen 
from these Figures, at distances $D\geq 400h^{-1}$Mpc this fraction 
of observed galaxies is strongly suppressed because of the radial 
selection effect. Note that this suppression is quite successfully 
fit by curves describing a selection function of the form
\be
f_{gal}(D)\propto D^2\exp[-(D/R_{sel})^{3/2}],\quad R_{sel}\approx 
190 h^{-1}{\rm Mpc}\,,
\label{sel}
\ee
where $D$ is a galaxy's radial distance and $R_{sel}$ is the 
selection scale \citep{BaEf:93}. These fits are also 
plotted in Fig. \ref{hst}. 

In some applications, like when we want to correct a measure of the
observed density to a measure of the true density, we would like to
use equation (\ref{sel}) to correct for the radial selection effects
after the fact.  An example of such a case is calculating a group's 
or cluster's true richness based upon the observed number of galaxies 
it contains (Sec. 6 \& 7). modified 

\begin{figure}
\centering
\epsfxsize=7cm
\hspace{0.5cm}
\epsfbox{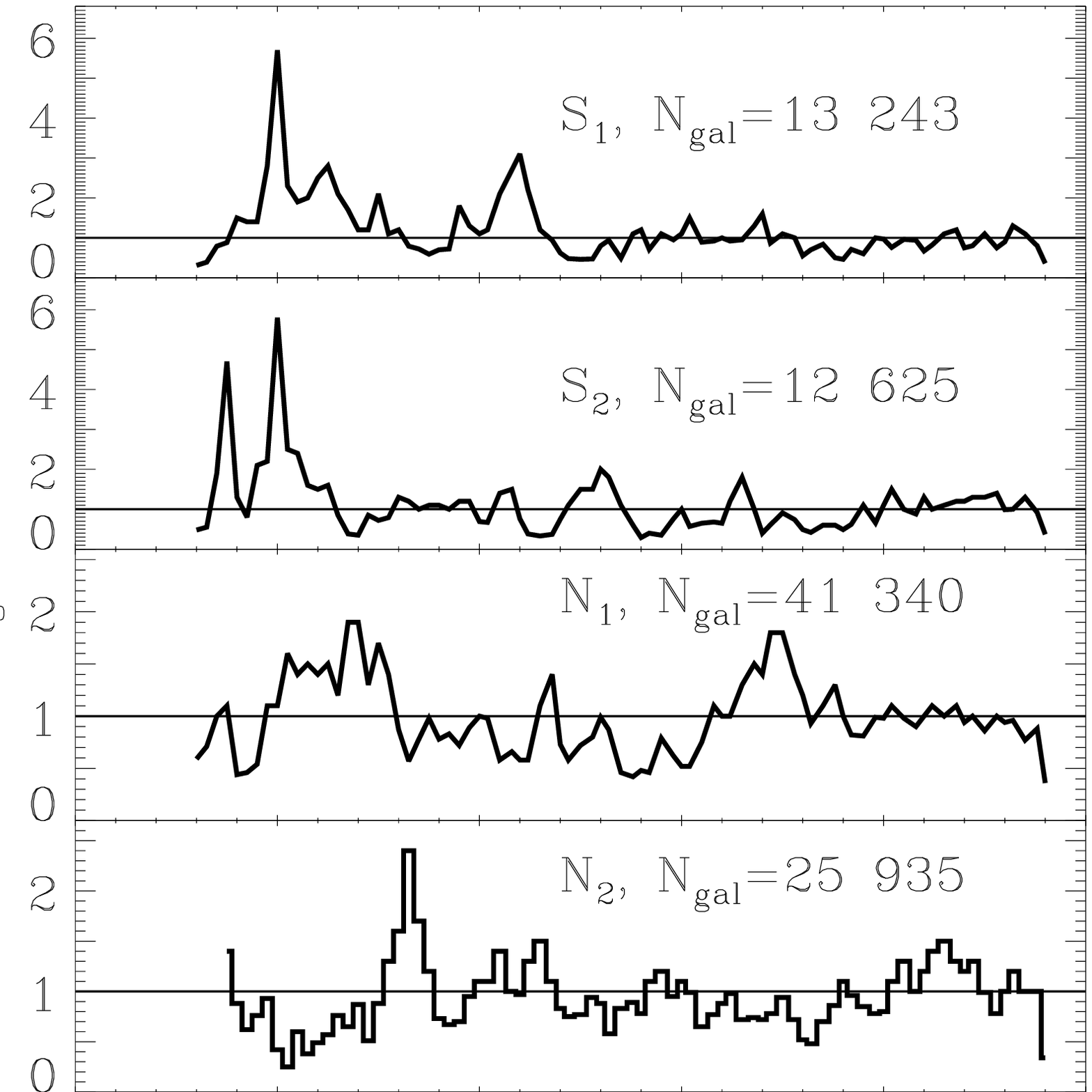}
\vspace{1.cm}
\caption{The normalized mean galaxy density in the four modified 
samples of SDSS DR1. 
} 
\label{rho}
\end{figure}

In other applications, however, like in searching for groups or
clusters in a magnitude-limited sample, we want to make a preemptive 
correction for the radial selection effects. For example, in a standard 
friends-of-friends percolation algorithm \citep[e.g.][]{HuGe:82}, 
this is done by adjusting the linking length as a function of radial 
distance. Here, instead, we employ the rather novel approach of 
adjusting the radial distances themselves as introduced and discussed 
in \cite{DoFo:01}. Hence, instead of the measured 
radial distance, we use a modified radial distance, $D_{md}$, where 
\be
D_{md}^3 = 2R_{sel}^3(1-[1+(D/R_{sel})^{3/2}]\exp[-(D/R_{sel}
)^{3/2}])\,.
\label{ra}
\ee 
The radial variations of the normalized number density of galaxies for
all samples from Figure \ref{hst} are plotted in Figure \ref{rho}. 
As is seen from this figure, the modified radial distances for the 
galaxies suppresses the very large-scale trends. On the other hand the
relative position of galaxies remain unchanged and the smaller scale 
random variations in the density are emphasized. The wedge diagramm 
of modified galaxy distribution for the sample $S_1$ is also plotted in 
Fig. \ref{wedge}.

This correction does not change distances at $D\leq R_{sel}$ and 
is more important for the more distant regions of our
samples ($D\geq 350 h^{-1}$Mpc), which contain only $\sim$20\% of all
galaxies. Thus, in the following analyses, we apply this correction
only to the separation of the high and low density regions in 
the deeper samples.  Of course, it cannot restore the lost 
information about the galaxy distribution in these regions, but 
it does help compensate for the strong drop in the observed galaxy 
density at these distances. It also allows one to apply the standard 
methods of investigation for the full catalogues with a depth of 
600$h^{-1}$Mpc.

\section{General characteristics of observed large scale structure}

To characterize the general properties of the large scale spatial galaxy 
distribution we use the Minimal Spanning Tree (MST) technique applied 
to both directly observed samples of galaxies and to samples corrected 
for the selection effect. 

In our analysis here, we consider the four fields, plotted in Fig. 
\ref{sky}, at the distance $D \leq 420 h^{-1}$Mpc with 
\be
N_{gal}=79\,183,\quad \langle n_{gal}\rangle\approx 10^{-2}
{\rm Mpc}^{-3}\,, 
\label{smp}
\ee
where $N_{gal}$ and $\langle n_{gal}\rangle$ are the total number 
of galaxies and the mean density of the samples. This sample contains 
$\approx$85\% of all galaxies with moderate impact from selection effects.
The numbers of galaxies in the separate fields are 
\renewcommand{\labelitemi}{--}
\begin{itemize}
\item $N_1$, the northern sample (35\,520 galaxies)
\item $N_2$, the northern sample (21\,983 galaxies)
\item $S_1$, the southern sample (11\,225 galaxies)
\item $S_2$, the southern sample (10\,455 galaxies)
\end{itemize}

\subsection{The MST technique}

The MST technique was first discussed by \cite{BaBhSo:85}
and by \cite{Weygaert1991}. 
The MST is a construct from graph theory, originally introduced 
by \cite{Kruskal:56} and \cite{Prim:57}, which has been widely applied in 
telecommunications and similar fields. It is a {\it unique network} 
associated with a given point sample and connects all points of the 
sample to a {\it tree} in a special and unique manner which minimizes 
the full length of the tree. Further definitions, examples, and
applications of this approach are discussed in \cite{BaBhSo:85} 
and \cite{Weygaert1991}. More references to the
mathematical results can also be found in \cite{Weygaert1991}.

One of earliest uses of MST approach in the study of large-scale 
structure was that of \cite{BhLi:88}, who successfully applied 
it to extract filamentary structures from the original CfA Redshift 
Survey. Its applications for the quantitative description of observed 
and simulated catalogues of galaxies were discussed in
\cite{DeDoTu:00,DoFo:00,DoFo:01}.

One of the most important features of the MST technique is 
generalization of the widely used ``friends--of--friends" approach. 
It allows one to separate all clusters, LSS elements, with a given 
linking length. In spite of the very complex shape of the clusters, 
the linking length defines for each two points the local overdensity 
bounding the clusters with a relation familiar from 
``friends--of--friends" algorithms \citep{HuGe:82}:
\be 
\delta_{thr} = 3/[4\pi\langle n_{gal}\rangle r_{lnk}^3]\,.
\label{rlnk}
\ee
Further on it allows one to obtain characteristics of each cluster and 
of the sample of clusters forming the LSS with a given overdensity. 
Further discrimination can be performed for a given threshold 
richness of individual elements. 

Here we will restrict our investigation to our results for the {\it 
probability distribution function of the MST edge lengths} $W_{MST}(l)$ 
and to the morphological description of individual clusters. The 
potential of the MST approach is not, however, exhausted by these 
applications. 

\subsection{Wall-like and filamentary structure elements}

With the MST technique we can demonstrate that the majority of 
galaxies are concentrated within wall--like structures and filaments 
which connect walls to the joint random network of the cosmic web. The 
internal structure of both walls and filaments is complex. Thus, 
wall--like structures incorporate some fraction of filaments and both 
walls and filaments incorporate high density galaxy groups and clouds. 
In particular, clusters of galaxies are usually situated within richer 
walls while groups of galaxies are embedded within filaments. In spite 
of this, the galaxy distribution can be described as a set of one, two 
and three dimensional Poisson--like distributions. Naturally, a one 
dimensional distribution is more typical for filaments while two and 
three dimensional ones are typical for walls and groups of galaxies, 
respectively. As was shown in \cite{Weygaert1991} \cite{BuDo:96},
a Poissonian distribution of galaxies within the LSS elements successfully
reproduces the observed 3D correlation function of galaxies. 

These result show that the probability distribution function of MST 
edge lengths (PDF MST), $W_{MST}(l)$, characterizes the geometry of 
the galaxy distribution.  For 1D and 2D Poissonian distributions,
typical for filaments and walls, $W_{MST}(l)$ is 
described by the following exponential and Rayleigh functions,
\begin{equation}
W_{MST}(l)=W_e(l)=\langle l\rangle^{-1}\exp(-l/\langle l\rangle)\,,
\label{W} 
\end{equation}
\[
W_{MST}(l)=W_R(l)=2l/\langle l^2\rangle\exp(-l^2/\langle l^2\rangle)\,. 
\]
These PDFs remain valid for any 1D and 2D distributions when 
the galaxy separation is small as compared with the curvature of 
the lines and surfaces. Comparison of measured and expected PDFs 
MST allows one to demonstrate the existence of these two types of 
structure elements and to make approximate estimates of their 
richness. Let us recall, however, that for the galaxy groups 
embedded within filaments, 2D and 3D Poissonian distributions are
observed and, so, in this case we cannot see a purely 1D distribution. 
For walls this effect is less important because it only distorts the
2D distribution typical for such LSS elements. 

\begin{figure}
\centering
\epsfxsize=7cm
\hspace{0.5cm}
\epsfbox{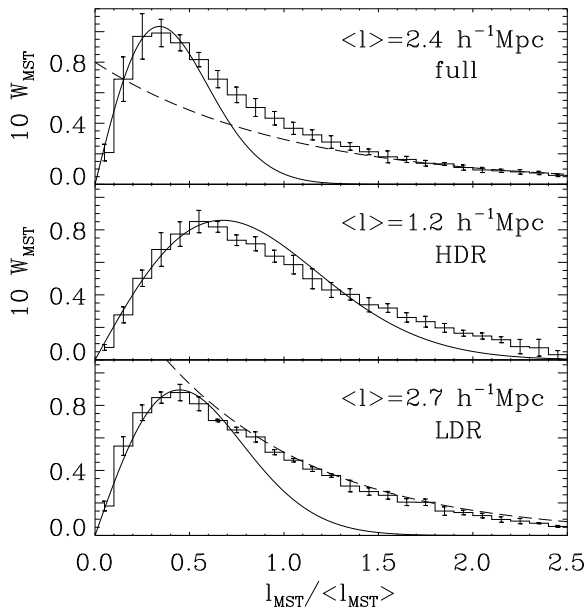}
\vspace{1.cm}
\caption{PDFs of MST edge lengths in redshift space averaged over 
four samples are plotted for the full sample (top panel), HDRs 
(middle panel) and LDRs (bottom panel). Rayleigh and exponential 
fits are plotted by thin solid and dashed lines.
} 
\label{mst}
\end{figure}

In Fig. \ref{mst} (top panel) we plott the $W_{MST}(l_{MST})$'s 
for the entire sample of 79,183 galaxies situated at distances $D\leq 
420 h^{-1}$Mpc where, as is seen from Fig. \ref{hst}, the impact 
of the selection effect is still moderate. The error bars show the 
scatter of measurements for the four subsamples. For each sample,
$N_{1},N_{2},S_{1},S_{2}$, we have
\be
\langle l_{MST}\rangle = 2.5,\,2.6,\,2.3,\,2.0\,h^{-1}{\rm Mpc}\,.
\label{lmst_i}
\ee
These variations demonstrate the differences in the sample properties 
(cosmic variance). 

Notice in Fig. \ref{mst} that the $W_{MST}(l_{MST})$ is well fit by a
superposition 
of Rayleigh (at $l_{MST}\leq \langle l_{MST}\rangle$) and exponential 
(at $l_{MST}\geq \langle l_{MST}\rangle$) functions. This confirms 
results discussed in Doroshkevich et al. (2000; 2001) in respect to 
the high degree of galaxy concentration within the population of high 
density rich wall--like structures and less rich filaments. However, 
as was noted in the same papers, with this approach the approximate 
separation of wall--like and filamentary structure elements can be 
performed only statistically. This is because the high density part of the 
PDF described by the Rayleigh function includes high density clouds 
situated in both filaments and walls. The exponential part of the PDF 
is related mainly to the filamentary component. 

\subsection{High and low density regions}

The methods for an approximate statistical decomposition of a
sample into subsamples of wall--like structures and filaments were 
proposed and tested in our previous publications 
\citep{DeDoTu:00,DoFo:00,DoFo:01}.  The first step is to make 
a rough discrimination between the high and low density regions 
(HDRs and LDRs).

Such discrimination can be easily performed for a given overdensity
contour bounding the clusters and a given threshold richness of
individual elements. Following \cite{DoFo:01}, in all 
four samples with $D\leq 420 h^{-1}$Mpc, wall--like high density 
regions (HDRs) were identified with clusters found for a threshold 
richness $N_{thr}=$ 40 and a threshold overdensity contour bounding 
the cluster equal to the mean density, $\delta_{thr}=1$. These 
samples, $N_{1},N_{2},S_{1},S_{2}$, of HDRs
contain 49\%, 47\%, 51\% and 47\% of
all galaxies. The samples of low density regions (LDRs), which are 
occupied mainly by filaments and poor groups of galaxies, are 
complementary to the HDRs in that the LDRs are simply the leftovers 
from the original total samples after the HDRs have been removed.

In Figure \ref{mst} (middle panel) the $W_{MST}(l)$ plotted for the 
HDRs is very similar to a Rayleigh function, thus confirming with 
this criterion the sheet-like nature of the observed galaxy distribution 
within the HDRs. As before, the error bars show the scatter of 
measurements for the four subsamples. For 90\% of objects we have
\be
W_{HDR}=(1\pm 0.18)~W_R,\quad l_{MST}\leq 1.65\langle l_{MST}\rangle\,,
\label{whdr}
\ee 
where $W_R$ is the Rayleigh function (\ref{W}). Larger difference between 
observed and expected PDFs for larger $l_{MST}$ indicates that the 
selected sample of HDRs includes some fraction of objects, $\sim$ 10\%,
which can be related to the filamentary component with the exponential PDF. 

For the LDRs, the $W_{MST}(l)$ is plotted in Fig. \ref{mst} (bottom 
panel). For small edge lengths, $l\leq \langle l_{MST}\rangle$, it 
also fits well to a Rayleigh function indicating that $\sim$ 60\% 
of LDR galaxies are concentrated within less massive 3D (elliptical) 
and 2D (sheet-like) clouds. This result confirms the strong disruption 
of filaments to a system of high density clouds. For larger edge 
lengths, however, the LDR $W_{MST}(l)$ appears to be closer to an 
exponential function, indicating that according to this criterion the
spatial distribution of the remaining $\sim$ 40\% of LDR galaxies is 
similar to a 1D Poissonian one which is typical for filamentary structures.

\begin{figure}
\centering
\epsfxsize=7cm
\hspace{0.5cm}
\epsfbox{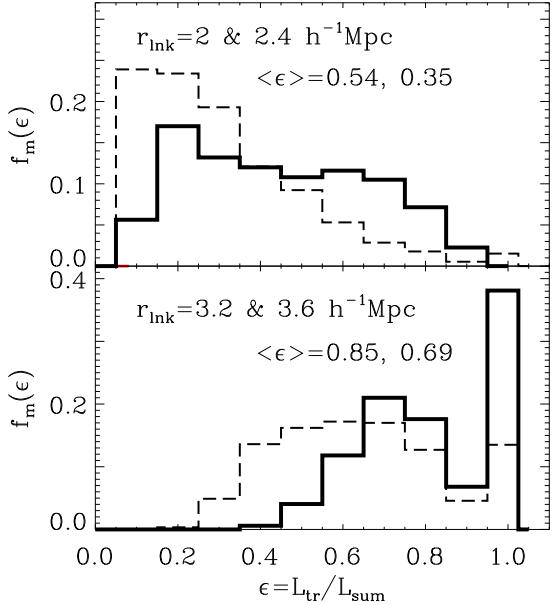}
\vspace{1.cm}
\caption{Mass functions of structure elements, $f_m(\epsilon)$, 
$\epsilon=L_{tr}/L_{sum}$ for the structure elements selected 
within HDRs (top panel, solid and dashed lines) 
and within LDRs (bottom panel, solid and dashed lines). 
} 
\label{eps}
\end{figure}

The mean edge lengths, $\langle l_{MST}\rangle$, found for HDRs and
LDRs in samples $N_{1},N_{2},S_{1},S_{2}$ are 
\be
\langle l_{MST}\rangle = 1.2,\,1.3,\,1.2,\,1.3\,h^{-1}{\rm Mpc}\,,
\label{lodr}
\ee
\be
\langle l_{MST}\rangle = 2.8,\,2.7,\,2.6,\,2.8\,h^{-1}{\rm Mpc}\,.
\label{ludr}
\ee
These values differ by about a factor of two from each other,
indicating that, as is seen from (\ref{rlnk}), the difference
in the mean density within 
HDRs and LDRs elements is about an order of magnitude. Of course, 
the volume averaged density of LDRs is still less.  

\subsection{Morphology of the structure elements}

Within so defined HDRs and LDRs themselves we can extract with the 
MST technique subsamples of structure elements for various threshold
overdensities. We can then suitably characterize the morphology of
each structure element by comparing the sum all edge lengths within
its full tree, $L_{sum}$, with the sum of all edge lengths within the
tree's trunk, $L_{tr}$, which is the longest path that can be traced
along the tree without re-tracing any steps. The ratio of these lengths 
\be
\epsilon =L_{tr}/L_{sum}\,.
\label{trs}
\ee
suitably characterizes the morphology of the LSS elements.

For filaments, we can expect that the lengths of the full tree 
and of the trunk are similar to each other, $\epsilon\sim 1$,  
whereas for clouds and walls these lengths are certainly very 
different and $\epsilon\leq 1$. This approach takes into account 
the internal structure of each element rather than the shape of 
the isodensity contour bounding it, and in this respect it is 
complementary to the Minkowski Functional technique
\cite[e.g.][]{ShGo:99,ShSa:02}.

However, even this method cannot discriminate between the wall--like
and 3D (elliptical) clouds and those rich filaments having many long 
branches for which again $\epsilon\leq$ 1. This means that both the 
PDF of this ratio, $W(\epsilon)$, and the corresponding mass function,
$f_m(\epsilon)$, are continuous functions and the morphology of
structure elements can be more suitably characterized by the degree 
of filamentarity and `wall-ness'.  This also means that we can only 
hope to distinguish statistical differences between the morphologies 
of structure elements in HDRs and the morphologies of structure elements 
in LDRs.

The selection of clusters within HDRs and LDRs was performed for 
two threshold linking lengths, $r_{lnk}=2. ~\&~2.4 h^{-1}$Mpc for 
HDRs, and $r_{lnk}=3.2 ~\&~3.6 h^{-1}$Mpc for LDRs. These values are 
larger than the mean edge lengths and characterize the LSS elements 
with intermediate richness when the measured difference between the 
walls and filaments is maximal. As was noted above, for lower linking 
lengths this method characterizes mainly the internal structure of 
the LSS elements while for large linking lengths filaments percolate 
and form the joint network with again $\epsilon\ll 1$. 

The distribution functions of the ratio, $W(\epsilon)$, are found to 
be close to Gaussian with $\langle\epsilon\rangle\approx 0.5 ~\&~ 
0.70$ for HDRs and LDRs, respectively. The mass functions, 
$f_m(\epsilon)$, plotted in Fig. \ref{eps} for the same linking 
lengths are shifted to the left (for HDRs) and to the right (for LDRs) with
respect to the middle point. Differences between these functions 
found for smaller and larger linking lengths illustrate the impact 
of the percolation process and disruption of the LSS elements. 

These results verify the objective nature of the differences in
the structure morphologies in HDRs and LDRs. 

\section{Statistical characteristics of filaments}

Theoretical characteristics of the LSS elements relate to the dominant 
dark matter (DM) component while the observed galaxy distribution 
relates to the luminous matter which represent only $\sim$ 3 -- 5\% 
of the mean density of the Universe. Spatial distribution of dark 
and luminous matter is strongly biased. None the less, some observed 
characteristics of filamentary components of the LSS can be compared 
with both available theoretical expectations and characteristics 
obtained for simulated DM distributions. 

The most interesting ones are the PDF of the linear density of 
filaments measured as the mass or number of objects per unit length 
of filament, $\Sigma_{fil}$. The other is the mean surface density 
of filaments 
$\sigma_{fil}$, defined as the mean number of filaments intersecting 
a unit area of arbitrary orientation. Both characteristics depend 
upon the threshold linking length, $r_{lnk}$, used for the filament 
selection and upon the threshold richness of filaments. However, both 
characteristics are independent from the small scale clustering of 
matter within filaments. 

Comparison of these characteristics of filaments for observed and 
simulated catalogues allows one to test the cosmological model used. 
However, the connection of quantitative characteristics of filaments 
with the initial power spectrum is complex and these measurements 
cannot yet be used for estimates of the power spectrum.

\subsection{Linear density of filaments}

As was found in DD99 and DD02, for the CDM--like initial power 
spectrum the PDF of the filament linear density can be written 
as follows: 
\be
N_{fil}~d\Sigma_{fil}\approx {1.5\over\langle\Sigma_{fil}\rangle}
\exp(-\sqrt{3\Sigma_{fil}/\langle\Sigma_{fil}\rangle})
~d\Sigma_{fil}\,.
\label{tfil}
\ee
However, poorer filaments cannot actually be selected in either
simulated or observed catalogues. Hence, even for high resolution 
simulations the measured PDF is well fitted by the relation
\be
N_{fil}\approx a_0~{\rm erf}^4[a_1(x-x_0)]\exp[-\sqrt{a_2(x-x_0)}] 
\label{sfil}
\ee
where $x=\Sigma_{fil}/\langle\Sigma_{fil}\rangle, x_0\approx 0.35, 
a_1\approx 2 - 2.5$ and $a_2\approx 30 - 40$ (DD02) and $\langle
\Sigma_{fil}\rangle\sim 3/r_{lnk}$. Here the cutoff of the PDF at 
$x=x_0$ reflects the limited resolution of simulated matter 
distribution. 

These results can be compared with ones obtained for for the DR1
and the mock catalogue \citep{CoHa:98} which allows one to estimate 
the impact of the selection effect for the measured linear density of 
filaments. In both the cases, the linear density of objects was 
measured by the ratio of the number of points and the length of the 
MST for each filaments of the sample. 

\begin{figure}
\centering
\vspace{0.2cm}
\epsfxsize=7cm
\hspace{0.5cm}
\epsfbox{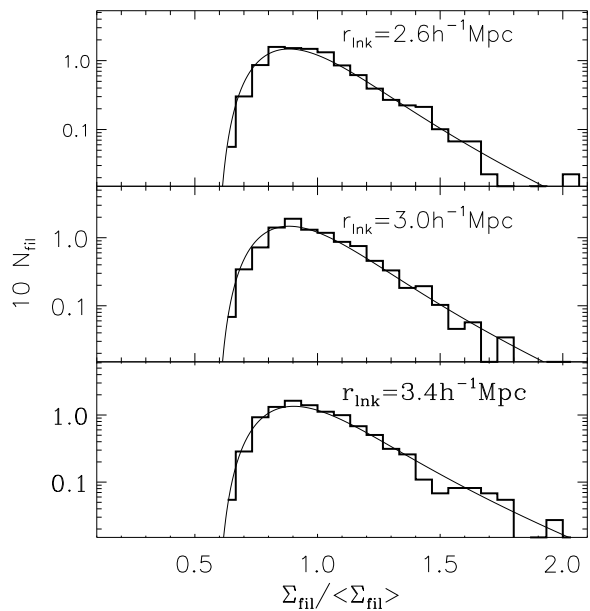}
\vspace{1.cm}
\caption{Distribution function, $N_{fil}$, for the linear density of 
'galaxies' along a tree for filaments selected in LDRs of the mock 
catalogues with three linking lengths. Fits (\ref{tfil}) are plotted 
by solid lines.
} 
\label{mfil}
\end{figure}

\begin{figure}
\centering
\vspace{0.2cm}
\epsfxsize=7cm
\hspace{0.5cm}
\epsfbox{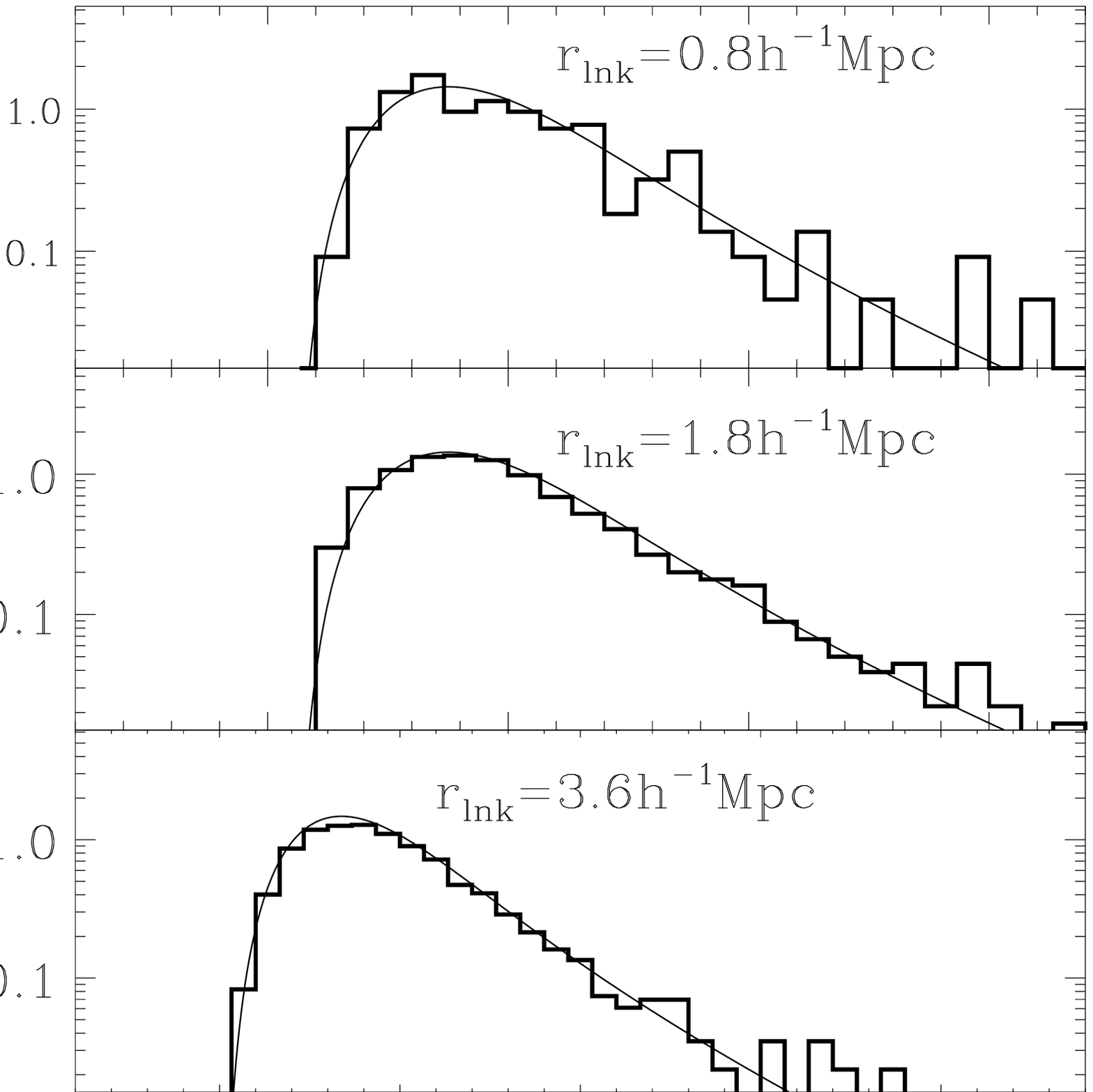}
\vspace{1.cm}
\caption{Distribution function, $N_{fil}$, for the linear density 
of galaxies along a tree for filaments selected in LDRs of the DR1 
with three linking lengths. Fits (\ref{tfil}) are plotted by solid 
lines.
} 
\label{ofil}
\end{figure}

For the mock catalogue the PDFs of the linear density of filaments
are plotted in Fig. \ref{mfil} for three linking length (with 
$\delta_{thr}\approx 1,~0.7~\&~0.5$). These PDFs are well fitted 
by expression (\ref{sfil}) for parameters $x_0\approx 0.6, a_1
\approx 2.5, a_2\approx 80 - 90$. For all linking lengths we get 
$\langle\Sigma_{fil}\rangle r_{lnk}= 2.25\pm 0.04$. For the 
observed DR1 catalogue the same PDFs are plotted in Fig. \ref{ofil} 
also for three linking length (with $\delta_{thr}\approx 50,~5~\&~ 
0.55$). They are well fitted by the same expression (\ref{sfil}) for 
parameters $x_0\approx 0.5, a_1\approx 2.5, a_2\approx 70 - 80$. 
For this sample we get $\langle\Sigma_{fil}\rangle r_{lnk}= 
2.2\pm 01$. 

These results show that, in all the cases, the measured PDFs are 
well fit by the same expression (\ref{sfil}) which coincides  
with the theoretically expected one (\ref{tfil}) at $\Sigma_{fil}
\geq\langle\Sigma_{fil}\rangle$. The mean linear density, $\langle
\Sigma_{fil}\rangle$, clearly depends upon the linking length 
used for filament selection. The selection effect increases the 
product $\langle\Sigma_{fil}\rangle r_{lnk}$ by $\sim$ 1.5 times 
as compared with results obtained for the DM simulations. 

These results show that the observed galaxy distribution nicely
represents the expected and simulated ones. The results also indicate 
that the general properties of filaments are consistent with a
CDM--like initial power spectrum. 

\subsection{Typical size of the filamentary network}

Due to the complex shape of the network of filaments spanning 
the LDRs any definition of the typical size of a network cell 
is mearly convenient. As was discussed in \cite{DoFo:01}, 
two definitions seem to be the most objective. One is the mean free 
path between filaments along a random straight line. The other is 
the mean distance between branch points of the tree along the 
trunk of selected filaments. The second definition tends to yield 
cell sizes that are typically a factor of 1.5 smaller than those 
yielded by the mean-free-path definition. 

Theoretical estimates of this size are uncertain because it strongly 
depends upon the sample of selected filaments (DD02). This means 
that this characteristic strongly depends upon the catalogue used. 
Moreover, filaments are connected to the network only for larger 
linking lengths; thus the typical measured cell size depends also 
upon the threshold linking length used. Hence, for the LCRS the mean 
free path between filaments with a variety of richness was estimated 
in \cite{DoFo:01} as $\sim 13-30 h^{-1}$Mpc. The mean 
distance between branch points of the tree along the trunk was 
estimated as $\approx 10 h^{-1}$Mpc and it rapidly increases with 
the linking length used owing the progressive percolation of 
filaments and formation of the joint LSS network. 

\begin{figure}
\centering
\vspace{0.2cm}
\epsfxsize=7cm
\hspace{0.5cm}
\epsfbox{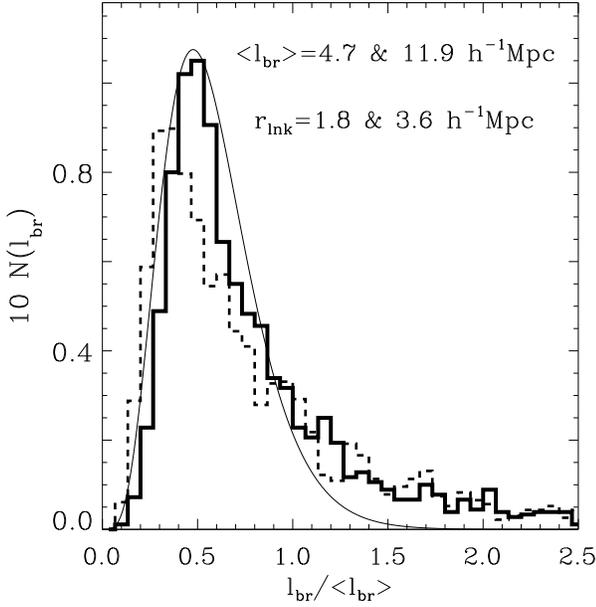}
\vspace{1.cm}
\caption{Distribution functions, $N$, for the distance between branch 
points along a trunk for filaments selected in LDRs (thick solid and 
dashed lines). Fit (\ref{lbr}) is plotted by thin solid line.
} 
\label{wbr}
\end{figure}

Here with a richer sample of filaments we can also estimate the 
PDFs of the cell sizes measured by the distance between branch 
points of the tree along the trunk. These PDFs, $N(l_{br})$, are 
plotted in Fig. \ref{wbr} for two linking lengths, $r_{lnk}=1.8 
~\&~3.6 h^{-1}$Mpc, which correspond to the threshold overdensities 
$\delta_{thr}=$ 0.66 \& 0.5. These PDFs are roughly fitted by 
expression 
\be
N(l_{br})\approx 270x^{4.5}\exp(-9.1 x),
\quad x=l_{br}/\langle l_{br}\rangle\,,
\label{lbr}
\ee
The measured mean distance between branch points, 
\[
\langle l_{br}\rangle\approx 4.7~\&~ 11.9 h^{-1}{\rm Mpc}\,.
\]
are close to those obtained in \cite{DoFo:01} and \cite{DoTu:96}
and those cited above. For smaller $r_{lnk}$ this estimate is 
decreased because of the domination of short filaments which are 
not yet connected to the network.  

\section{Parameters of the wall--like structure elements}

The statistical characteristics of observed walls were first measured
using the LCRS and DURS \citep{DoFo:00,DoFo:01}. The rich
sample of walls extracted from the SDSS DR1, however, permits more
refined estimates of these characteristics. As was discussed in Sec. 
3.2 walls dominate the HDRs, and thus these subsamples of galaxies 
can be used to estimate the wall properties.

The expected characteristics of walls and methods of their measurement
were discussed in \cite{DeDoTu:00} so here we will only
briefly reproduce the main definitions. It is important that these
characteristics can be measured independently in radial and transverse
directions, which reveals the strong influence of the velocity
dispersion on other wall characteristics.

\subsection{Main wall characteristics}

Main characteristic of walls is their mean dimensionless surface
density, $\langle q_w\rangle$, measured by the number of galaxies per
Mpc$^2$ and normalized by the mean density of galaxies multiplied by
a coherent length of the initial velocity field (DD99; DD02)
\be
l_v\approx 33 h^{-1}{\rm Mpc}~(0.2/\Gamma),\quad \Gamma = \Omega_m h\,,
\label{lv}
\ee
where $\Omega_m$ is the mean matter density of the Universe. 
For Gaussian initial perturbations, the expected probability 
distribution function (PDF) of the surface density is 
\be
N_{th}(q_w)= {1\over\sqrt{2\pi}}{1\over\tau_m\sqrt{q_w}}\exp\left(
-{q_w\over 8\tau_m^2}\right){\rm erf}\left(\sqrt{q_w\over 
8\tau_m^2}\right)\,,
\label{wq}
\ee
\[
\langle q_{th}\rangle = 8(0.5+1/\pi)\tau_m^2\approx 6.55\tau_m^2\,.
\]
This relation links the mean surface density of walls with the 
dimensionless amplitude of perturbations, $\tau_m$,
\be
\tau_m=\sqrt{\langle q_w\rangle /6.55}\,, 
\label{taum}
\ee
which can be compared with those measured by other methods (DD02, 
Sec. 8.2). 

Other important characteristics of walls are the mean velocity
dispersion of galaxies within walls, $\langle w_w \rangle$, the mean
separation between walls, $\langle D_{sep}\rangle$, the mean
overdensity, $\langle\delta\rangle$, and the mean thickness of walls,
$\langle h\rangle$. The mean velocity dispersion of galaxies, $\langle
w_w \rangle$, can be measured in radial direction only whereas other
wall characteristics can be measured both radially and along
transverse arcs. Comparison of the wall thickness and the overdensity,
$\langle h\rangle$ and $\langle\delta\rangle$, measured in transverse
($t$) and radial ($r$) directions, illustrates the influence of the
velocity dispersion of galaxies on the observed wall thickness.

The velocity dispersion of galaxies within a wall $w_w$ can be related
to the radial thickness of the wall by this relation \citep{DeDoTu:00}:
\be
h_r = \sqrt{12} H_0^{-1} w_w\,.
\label{ww_to_hr}
\ee

For a relaxed, gravitationally confined wall, the measured wall
overdensity, surface density, and the velocity dispersion are linked
by the condition of static equilibrium.  Consider a wall as a slab in
static equilibrium, and this slab has a nonhomogeneous matter
distribution across it.  We can then write the condition of static
equilibrium as follows:
\be
w_w^2 = {\pi G\mu^2\over\langle\rho\rangle \delta}\Theta_\Phi = 
{3\over 8}{\Omega_m\over\delta}(H_0l_vq_w)^2\Theta_\Phi\,,
\label{phi}
\ee
Here $\mu=\langle\rho\rangle l_v q_w$ is the mass surface density of
the wall and the factor $\Theta_\Phi\sim$ 1 describes the
nonhomogeneity of the matter distribution across the
slab. Unfortunately, for these estimates we can only use the velocity
dispersion and overdensity measured for radial and transverse
directions. Hence, the final result cannot be averaged
over the samples of walls.

\subsection{Measurement of the wall characteristics}

The characteristics of the walls can be measured with the two
parameter core--sampling approach \citep{DoTu:96} applied
to the subsample of galaxies selected within HDRs. With this method,
all galaxies of the sample are distributed within a set of radial
cores with a given angular size, $\theta_c$, or within a set of
cylindrical cores oriented along arcs of right ascension with a size
$d_c$. All galaxies are projected on the core axis and collected to a
set of one-dimensional clusters with a linking length, $l_{link}$. The
one-dimensional clusters with richnesses greater than some threshold
richness, $N_{min}$, are then used as the required sample of walls
within a sampling core.

Both the random intersection of core and walls and the nonhomogeneous 
galaxy distribution within walls lead to significant random scatter 
of measured wall characteristics. The influence of these factors 
cannot be eliminated, but it can be minimized for an optimal range 
of parameters $\theta_c$, $d_c$, $l_{link}$ and $N_{min}$. Results 
discussed below are averaged over the optimal range of these 
parameters. 

\begin{table*}
\begin{minipage}{160mm}
\begin{center}
\caption{Wall properties in observed and simulated catalogues}
\label{tbl2}
\begin{tabular}{lrc ccc ccc c} 
sample&$N_{gal}$&$\langle q_w\rangle/\Gamma$&$\tau_m/\sqrt{\Gamma}$&
$\langle\delta_r\rangle$&$\langle\delta_t\rangle$&$\langle h_r\rangle$&
$\langle h_t\rangle$&$\langle w_w \rangle$&$\langle D_{sep}\rangle$\cr
  &  &  &  & &$h^{-1}$Mpc&$h^{-1}$Mpc& km/s&$h^{-1}$Mpc \cr
\hline
\multicolumn{9}{c}{radial cores}\cr 
N1&41\,217&$2.01\pm 0.21$&$0.55\pm 0.03$&1.3&-&$10.2\pm 1.9$&-&$295\pm 55$&$69\pm   14$\cr 
N2&25\,935&$1.61\pm 0.16$&$0.49\pm 0.03$&1.1&-&$10.5\pm 2.0$&-&$303\pm 57$&$78\pm   17$\cr 
S1&13\,215&$2.21\pm 0.19$&$0.58\pm 0.03$&1.4&-&$10.5\pm 1.7$&-&$303\pm 50$&$67\pm   14$\cr 
S2&12\,585&$1.79\pm 0.33$&$0.51\pm 0.04$&1.3&-&$~~9.0\pm 1.8$&-&$260\pm 50$&$83\pm  20$\cr 
\hline
\multicolumn{9}{c}{transverse cores for the SDSS DR1}\cr 
N1&16\,883&$2.47\pm 0.51$&$0.61\pm 0.07$&- &3.9&- &$4.3\pm 0.8$&   -   &$65\pm  11$\cr
S1&13\,215&$2.29\pm 0.64$&$0.58\pm 0.08$&- &3.9&- &$4.0\pm 0.8$&   -   &$58\pm  11$\cr
\hline
\multicolumn{9}{c}{observed samples}\cr 
SDSS (radial)&92\,952&$1.91\pm 0.32$&$0.53\pm 0.05$&1.3&-&$10.1\pm 2.0$&-&$291\pm 56$&$74\pm 17$\cr 
SDSS (transverse)  &29\,311&$2.42\pm 0.67$&$0.60\pm 0.08$&- &3.5&- &$4.9\pm 1.3$&   -   &$64\pm  14$\cr
      LCRS&16\,756&$2.51\pm 0.9$&$0.62\pm 0.10$&3.0&7.4&$~~8.6\pm 0.8$&$2.8\pm 0.7$&$247\pm 48$&$60\pm 10$\cr
      DURS&2\,500&$2.23\pm 0.6$&$0.58\pm 0.08$&1.7&6.5&$~~9.7\pm 1.8$&$4.9\pm 1.2$&$280\pm 52$&$44\pm 10$\cr
\hline
\multicolumn{9}{c}{mock catalogues in real and redshift spaces for the model with $\Gamma=0.2$}\cr 
redshift&98\,828&$2.7\pm 0.5$&$0.63\pm 0.06$&1.8&3.8&$11.8\pm 2.1$&$6.5\pm 1.4$&$338\pm 65$&$50\pm  10$\cr
real    &98\,828&$2.1\pm 0.4$&$0.57\pm 0.06$&4.3&4.6&$~~4.8\pm 1.0$&$4.2\pm 1.0$&$305\pm 47$&$50\pm 10$\cr
\hline
\multicolumn{9}{c}{DM catalogue in real space for the model with $\Gamma=0.2$}\cr 
real &$7.1\cdot 10^6$&$2.5\pm 0.4$&$0.63\pm 0.04$&2.7&   &$~~4.9\pm 0.5$&$     $&$245\pm 30$&$52\pm  5$\cr
\hline
\end{tabular}
\end{center}
\end{minipage}
\end{table*}

For the measurement of wall characteristics in the radial direction 
four samples of HDRs galaxies were used in each field of the DR1 
catalogue. One of these samples was selected as was discussed in Sec. 
3.3, three other samples were selected from the catalogues 
{\it already corrected for radial selection effects} 
(Sec.~2.1) with the same threshold 
overdensity $\delta_{thr}=1$ and for HDRs containing $\sim$ 43\%, 
50\%, and 56\% of all galaxies in the field. In all the cases, the 
wall parameters were measured in real space for the selected samples 
of the HDRs. The mean wall properties were averaged over four radial 
core sizes ($\theta_c = 2^\circ, 2.25^\circ, 2.5^\circ$ and 
$2.75^\circ$) and for six core-sampling linking lengths ($2 
h^{-1}$Mpc$\leq l_{link}\leq 4.5 h^{-1}$Mpc). Final averaging was 
performed over all sixteen samples and over all $\theta_c$ and 
$l_{link}$. 

Due to the complex shape of the fields $S_2$ and $N_2$, the measurements 
of wall characteristics in the transverse direction were performed 
for the fields $S_1$ and $N_1$ only. The mean wall properties were 
averaged over four core diameters ($d_c=$ 6.0, 6.5, 7.0, and 7.5
$h^{-1}$Mpc) and five core-sampling linking lengths ($2 h^{-1}$Mpc
$\leq l_{link}\leq 4. h^{-1}$Mpc). 

\subsection{Measured characteristics of walls}

The mean radial and tranverse wall properties for all fields are 
listed separately in Table~1. Characteristics obtained by averaging 
over all samples are compared with those from the DURS and LCRS and 
with those from mock catalogues simulating the SDSS EDR \citep{CoHa:98}. 
Both DM simulation and mock catalogues are prepared for the 
$\Lambda$CDM cosmological model with $\Omega_m=0.3$, $\Omega_\Lambda
=0.7$ and with the amplitude of perturbations $\sigma_8=1.05$ that 
exceed the now accepted value $\sigma_8=0.9\pm 0.1$ \citep{Sper:03}.
This excess of the amplitude increases both the measured 
$\tau_m$ and $\langle q\rangle$ by $\sim$ 10\% and $\sim$ 20\%, 
respectively. 

The richness and geometry of these catalogues are strongly 
different. Thus, DURS is an actual 3D catalogue but it contains 
$\sim$ 2 500 galaxies at the distance $D\leq 250 h^{-1}$ Mpc and 
its representativity is strongly limited. The LCRS include $\sim$ 
21\,000 galaxies at the distance $D\leq 450 h^{-1}$ Mpc but they 
are distributed within six thin slices that again distorts the 
measured wall characteristics. Moreover, both catalogues include a
small number of walls with large scatter in richness. This leads 
to a significant scatter of measured characteristics of walls for 
these catalogues which is a manifestation of well known cosmic 
variance. Only with an actually representative catalogue such as the 
the SDSS can this effect can be suppressed. 

The difference between the mean wall surface densities measured 
for $\sim$ 15--20\% of samples reflects real variations in
wall properties for a limited portion of the 
samples. However, the scatters of mean values listed in Table 1
partially include the dispersions depending on the shape of their
distribution functions. The actual scatter of the mean characteristics 
of walls averaged over all samples listed in Table 1 is also 
$\leq$ 10--12\%.

The amplitude of initial perturbations characterized by values
$\tau_m$ for the richer sample measured in the radial direction 
is 
\be
\tau\approx (0.53\pm 0.05)\sqrt{\Gamma} = (0.24\pm 0.02)
\sqrt{\Gamma\over 0.2}\,.
\label{ttau}
\ee
This is quite consistent with estimates found for simulations. 
Differences between this value and $\tau_m$ measured in the transverse 
direction for the DR1, the LCRS and the DURS demonstrate the impact of 
the representativity of the catalogue used. 

The difference between the wall thickness measured in the radial 
and transverse directions, $h_r$ and $h_t$, indicates that, 
along a short axis, the walls are gravitationally confined 
stationary objects. Just as with the `Finger of God' effect 
for clusters of galaxies, this difference characterizes the 
gravitational potential of compressed DM rather than the 
actual wall thickness. The same effect is seen as a difference 
between the wall overdensities measured in radial and transverse 
directions. 

The difference between the wall thicknesses is compared with 
the velocity dispersions of galaxies within the walls, $\langle 
w_w\rangle$. Clusters of galaxies with large velocity dispersions 
incorporated in walls also increase the measured velocity 
dispersion. The correlation between the wall surface density 
and the velocity dispersion confirms the relaxation of matter 
within walls. This relaxation is probably accelerated due to 
strong small scale clustering of matter within walls. 
 
\begin{figure}
\centering
\epsfxsize=7cm
\hspace{0.5cm}
\epsfbox{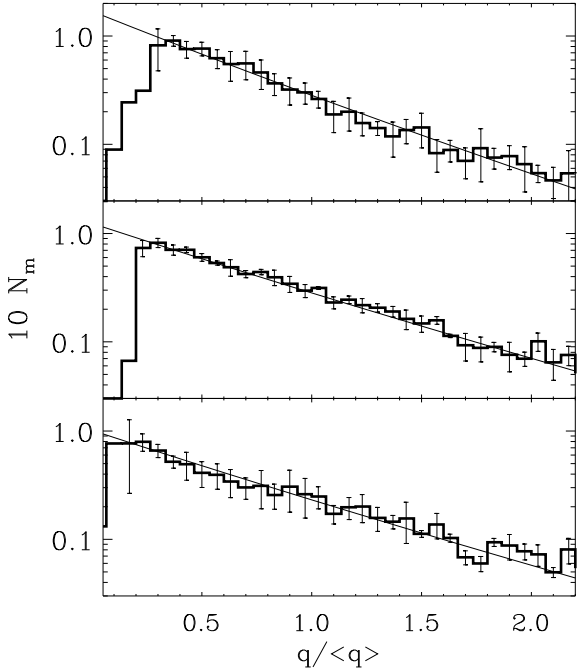}
\vspace{1.cm}
\caption{The PDFs of dimensionless surface density of walls, 
$N_m(q/\langle q\rangle)$, for walls selected in the DM simulation 
(top panel) and the mock catalogue in real (middle panel) and 
redshift (bottom panel) spaces. Fits (\ref{wq}) are plotted by 
solid lines.
} 
\label{qsim}
\end{figure}

\begin{figure}
\centering
\epsfxsize=7cm
\hspace{0.5cm}
\epsfbox{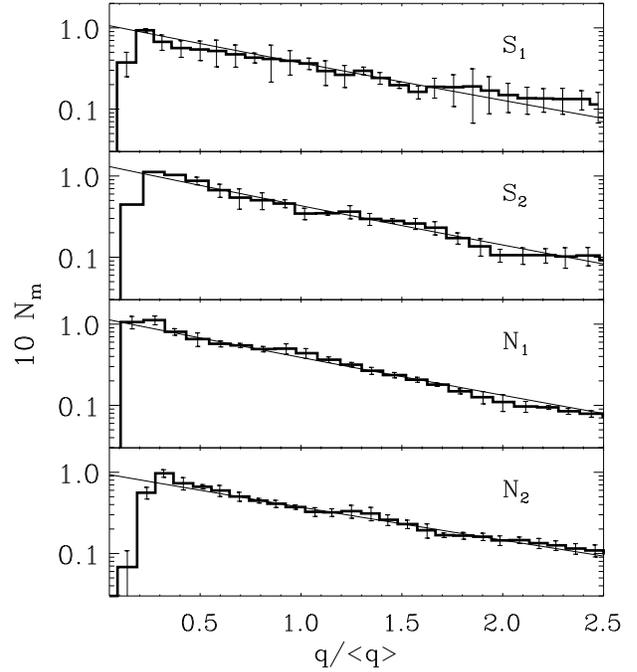}
\vspace{1cm}
\caption{The PDFs of observed dimensionless surface density 
of walls, $N_m(q/\langle q\rangle)$, for walls selected 
in four regions of the SDSS galaxy catalogues. Theoretically 
expected fits are plotted by solid lines.
} 
\label{qobs}
\end{figure}

The measured PDFs of the surface density of walls are plotted 
in Figs. \ref{qsim} and \ref{qobs} for the simulated DM distribution, 
mock catalogues in real and redshift spaces, and four observed 
samples of the SDSS DR1. These are nicely fit by the expected 
expression (\ref{wq}). Thus, for simulated samples we have, 
respectively,  
\be
N_m=(1\pm 0.1)N_{th},\quad \langle q\rangle/\langle q_{th}\rangle=
0.78, 0.87, 0.87\,,
\label{esim}
\ee
where $N_{th}$ and $\langle q_{th}\rangle$ are given by (\ref{wq}). 
For the two southern samples of the SDSS DR1 we get:
\be
N_m=(1.1\pm 0.2)N_{th},\quad \langle q\rangle/\langle q_{th}\rangle=
1.07\,,
\label{pobs1}
\ee
and for the two northern samples of the SDSS DR1 we get:
\be
N_m=(1.\pm 0.1)N_{th},\quad \langle q\rangle/\langle q_{th}\rangle=
0.97\,,
\label{pobs2}
\ee
These results verify that, the observed walls represent 
recently formed Zel'dovich' pancakes.

\begin{figure}
\centering
\epsfxsize=7cm
\hspace{0.5cm}
\epsfbox{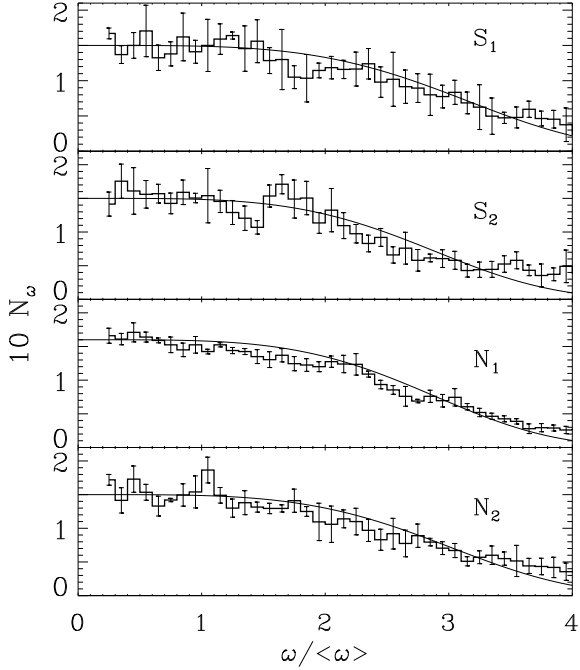}
\vspace{1.cm}
\caption{The PDFs, $N_{w}(w/\langle w\rangle)$, of reduced 
velocity dispersion within wall (\ref{omw}) 
for walls selected in four regions of the SDSS galaxy catalogues. 
Gaussian fits are plotted by solid lines.
} 
\label{wobs}
\end{figure}

Using the measured mean wall overdensity in the transverse direction 
listed in Table 1 we have for the parameter 
$\Theta_\Phi$ introduced in equation~(\ref{phi})
\be
\Theta_\Phi\approx {\langle\delta\rangle\over 3}{0.3\over\Omega_m}
\approx 1.1\,,
\label{pphi}
\ee
which is also consistent with the expected values for relaxed 
and stationary walls. 

As was proposed in \cite{DeDoTu:00} we can discriminate
between systematic variations in the measured velocity dispersion 
due to regular variations in the surface density along the walls 
(Fig. \ref{qobs}) and the random variations in the velocity 
dispersion which integrates the evolutionary history of each wall. 
Indeed, along a shorter axis, for gravitationally bound and relaxed
walls we can expect that 
\[
w_w^2\propto q^2/\delta\propto \delta^{\gamma-1},
\quad w_w\propto q^{1-1/\gamma}\,.
\]
Here we assume that the distribution of DM component and galaxies 
can be approximately described by the polytropic equation of state 
with the power index $\gamma\approx$ 5/3 -- 2. 
\cite{DeDoTu:03} suggest for 
consideration a reduced velocity dispersion, $\omega_w$, 
\be 
\omega_w= |\ln(w_w q_w^{-p_w})|,\quad p_w\approx 1-1/\gamma
\approx 0.5\,,
\label{omw}
\ee
corrected for variations of the wall thickness. For this 
function the systematic variations of $w_w$ are essentially 
suppressed and, in most respects, it is similar to the entropy 
of compressed matter. It integrates the action of random 
factors in the course of wall formation. Hence, for this 
function the Gaussian PDF, $N_\omega$, can be expected. Indeed, 
this PDF plotted in Fig. \ref{wobs} for four samples of the 
SDSS DR1 is quite similar to a Gaussian function with a standard 
deviation $\sigma_\omega\approx 2.25\langle \omega\rangle$. These 
results show a large scatter of evolutionary histories of observed 
walls. 

\begin{figure}
\centering
\epsfxsize=7cm
\hspace{0.5cm}
\epsfbox{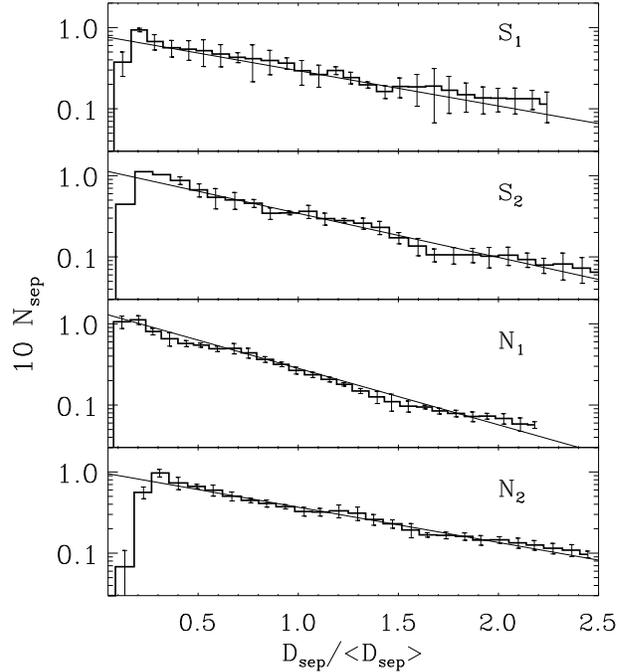}
\vspace{1.cm}
\caption{The PDFs, $N_{sep}(D_{sep}/\langle D_{sep}\rangle)$, 
of observed wall separations for walls selected in four regions of 
the SDSS galaxy catalogues. Theoretically expected for the 
Possonian distribution exponential fits are plotted by solid 
lines.
} 
\label{sobs}
\end{figure}

Note that, for all the samples listed in Table~1, the mean wall
separation, $\langle D_{sep}\rangle$, is close to twice that of the
coherent length of the initial velocity field,
\be
\langle D_{sep}\rangle\approx 2 l_v\,, 
\label{sep}
\ee
for the low density cosmological models with $\Gamma\approx 0.2$ 
(\ref{lv}). These results coincide with the estimates 
of the matter fraction, $\sim$ 50\%, accumulated within walls. 
Due to the large separation of walls, the correlations of their 
positions is small and a random 1D Poissonian PDF of the separation 
can be expected. These PDFs are plotted in Fig. \ref{sobs} together 
with the exponential fits. 

Finally, we would like to draw attention to the fact that all measured
properties of these walls are quite consistent with a CDM--like initial
power spectrum and Gaussian distribution of perturbations.

\section{Possible rich clusters of galaxies}

The SDSS DR1 also contains a number of galaxy complexes of
various richnesses which can be extracted by means of the MST
technique. Due to the large velocity dispersion of galaxies within
clusters and the strong `Finger of God' effect, this extraction must
be performed using different threshold linking lengths in the radial 
($l_r$) and in the transverse ($l_t$) directions.  This is not
unlike how group catalogs are extracted from redshift surveys
using conventional `friends-of-friends' algorithms
\citep{HuGe:82,Tu:00}.

\begin{figure*}     
\begin{minipage}{160mm}     
\epsfxsize=15.cm     
\hspace{1.cm}     
\epsfbox{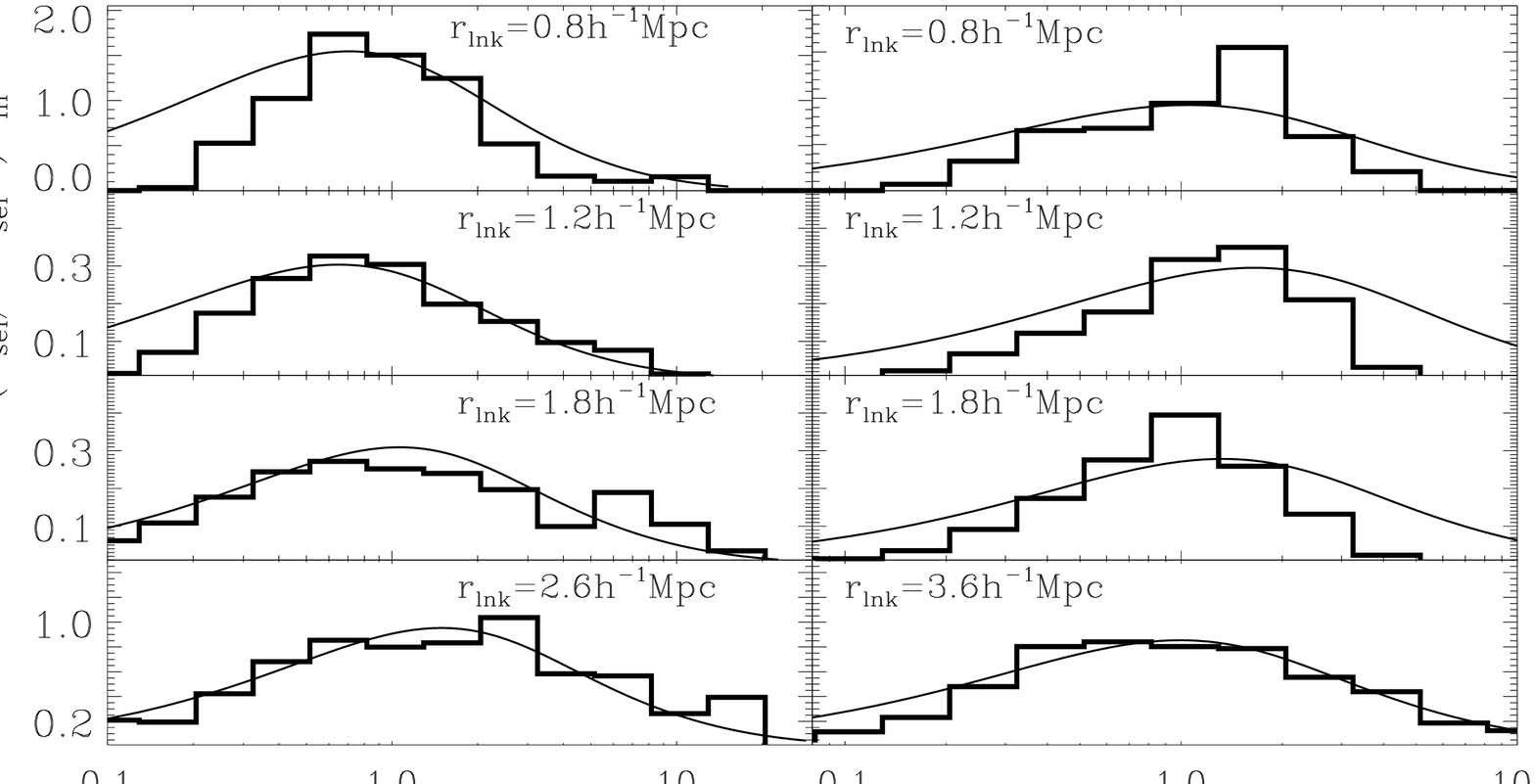}  
\vspace{1.2cm}     
\caption{Mass functions of galaxy clouds, $N_m\cdot N_{mem}/\langle      
N_{mem}\rangle)$, selected in HDRs (left panels) and LDRs (right 
panels) for four threshold linking lengths. Theoretical fit (\ref{nm1}) 
for relaxed clouds ($r_{lnk} = 0.8 \& 1.2 h^{-1}$~Mpc) and fit 
(\ref{nm2}) for unrelaxed clouds ($r_{lnk}\geq 1.8h^{-1}$~Mpc) 
are plotted by solid lines. 
}     
\end{minipage} 
\label{mass}    
\end{figure*}  

We performed this cluster-finding in two major steps.  First, we
projected the observed samples onto a sphere of radius $R=100 
h^{-1}$Mpc with a random scatter $\pm 0.5 h^{-1}$Mpc and extracted 
a set of candidate clusters from this catalog using a linking 
length of $r_t=0.3h^{-1}$Mpc. Second, we applied a radial
linking length of $r_r=3 h^{-1}$Mpc to these candidate clusters
using their real 3D coordinates. In this second step, we also 
employed the threshold richness, $N_{mem}=10$, for our final 
samples of possible rich complexes. Having extracted these 
probable rich ``clusters", we calculated a distance-independent 
measure of their richnesses by correcting their observed richnesses 
$N_{mem}$ for radial selection effects using equation (1); we call 
this corrected richness $N_{sel}$. Further discrimination of the 
"clusters" can be performed using their size in transverse and 
radial directions as well as other characteristics of selected galaxies. 

For the threshold parameters used 20, 12, 43 and 10 possible rich 
clusters  were selected from the $S_1, S_2, N_1\,\&\,N_2$ samples, 
respectively. The majority of these "clusters" are embedded within 
richer walls.  Let us remember that these are {\em possible\/} rich 
clusters of galaxies and to confirm that they are physical potential 
wells, it is necessary to check for diffuse x-ray emission. However, 
comparison with the list of NORAS survey \citep{Bo:00} 
shows that only $\sim$ 32 of them can be roughly related to the 
x-ray sources. This means that the method of cluster identification 
must be essentially improved. 

\section{Mass function of the structure elements}

The richness of the SDSS DR1 allows one to extract several different
sets of high density clouds and structure elements with various
overdensities within the HDRs and LDRs and to find their
mass function. These results can be directly compared with the
theoretical expectations of DD02. However, the  richness of 
the SDSS DR1 does not yet allow one to estimate quantitatively the 
divergence between the expected and observed mass functions 
plotted in Fig. \ref{mass}.

Two samples of high density galaxy groups and two samples of
unrelaxed structure elements -- walls and filaments -- were 
selected separately within HDRs and LDRs introduced in Sec. 3.3 
for a threshold richness $N_{mem}\geq$ 5. Since the velocity 
dispersions in groups are expected to be much smaller than those in rich 
clusters, we select these samples of structure elements using 
the simpler method described in Sec.~4 rather than the two-step 
approach described in Sec.~6.  The richness of each cluster was 
corrected for radial selection effects using the selection function 
introduced in Sec. 2.1. 

The main properties of the selected clouds are listed in Table~2, where 
$r_{lnk}$ and $\langle\delta\rangle$ are, respectively, the threshold 
linking length and the mean overdensity of clouds, $f_{gal}$ is the 
fraction of galaxies from the total (combined HDR+LDR) sample of 
galaxies within the selected clouds, $N_{cl}$ is the number of clouds while
$\langle M\rangle$ and $\langle M_{sel}\rangle$ are the observed  
and corrected for the selection effect mean richness of individual 
clouds.

Results listed in Table 2 illustrate the influence of environment on
the properties of high density clouds. In particular, in spite of the
approximately equal number of galaxies in HDRs and LDRs, $\sim$85\% 
and $\sim$70\% of the high density clouds selected with linking lengths 
$r_{lnk}=0.8~\&~ 1.2 h^{-1}$Mpc are situated within the HDRs and
accumulate $\sim$90\% and $\sim$80\% of galaxies. At linking length 
$r_{lnk}=1.8 h^{-1}$Mpc the numbers of clouds selected within HDRs and 
LDRs are comparable but again $\sim$80\% of galaxies related to these 
clouds are concentrated within HDRs. At the largest linking lengths 
listed in Table 2 essential differences are seen only for the mean 
masses and number of structure elements selected within HDRs and 
LDRs. These differences are enhanced by the influence of the selection 
effect which is stronger for the LDRs. 

The mass functions for these samples are plotted in Fig. \ref{mass}. 
As was shown in DD02, in Zel'dovich theory and for the WDM initial 
power spectrum the dark matter mass function of structure elements 
is independent of their shapes and, at small redshifts, it can be 
approximated by the functions
\begin{equation}
xN(x)dx = 12.5\kappa_{ZA}x^{2/3}\exp(-3.7x^{1/3})~{\rm erf}
(x^{2/3})dx\,,
\label{nm1}
\end{equation}
\begin{equation}
xN(x)dx = 8.\kappa_{ZA}x^{1/2}\exp(-3.1x^{1/3})~{\rm erf}
(x^{3/4})dx\,.
\label{nm2}
\end{equation}
\[x=\mu_{ZA}{M\over \langle M\rangle}\,,
\]
The expression (\ref{nm1}) relates to clouds which have become
essentially relaxed and static by $z=0$, and the expression
(\ref{nm2}) relates to richer, unrelaxed filaments and walls which 
are still in the process of collapse. Here, $\kappa_{ZA}\sim$ 1.5 
-- 4 and $\mu_{ZA}\sim$ 0.8 -- 1.3 are fit parameters which take 
into account the incompleteness of selected samples of clouds for 
small and large richnesses; this incompleteness changes both the 
amplitude and mean mass of the measured clouds. Comparison with 
simulations (DD02) has shown that these relations fit reasonably 
well to the mass distribution of DM structure elements.

\begin{table}
\caption{Parameters of groups of galaxies selected 
in HDRs and LDRs after correction for the selection effect.} 
\begin{center}
\begin{tabular}{crl rrr} 
\hline
$r_{lnk}h^{-1}$Mpc&$\langle\delta\rangle$&$f_{gal}$&$N_{cl}$
&$\langle M\rangle$&$\langle M_{sel}\rangle$\\
\hline
&&HDR&&\\
0.8&140   &0.1  &1\,193& 6.8&12\\
1.2& 40   &0.24 &1\,961& 9.5&25\\
1.8& 10   &0.39 &1\,731&17.7&61\\
2.6&  3   &0.47 &   614&61.5&228\\
\hline
&&LDR&&\\
0.8&220&0.014&   220&4.8 &14\\
1.2& 61&0.05 &   817&5.3 &18\\
1.8& 16&0.15 &1\,805&6.4 &36\\
3.6&  3&0.36 &2\,295&12.5&106\\
\hline
\end{tabular}
\end{center}
\end{table}

These relations are similar to the mass function from the Press--
Schechter formalism,
\begin{equation}
xN_{PS}(x)dx = {8 \kappa_{PS}\over 45\sqrt{\pi}}~\xi^{1/6}
\exp(-\xi^{1/3})dx\,,
\label{ps}
\end{equation}
\[
\xi=1.785\mu_{ps} x=1.785\mu_{ps} M/\langle M\rangle\,,
\]
despite the fact that they use different assumptions about the 
process of cloud formation and the shape of the formed clouds.
Here again the fitting parameters $\kappa_{PS}$ and $\mu_{PS}$ 
take into account the incompleteness of measured sample. However, 
this expression relates to the CDM-like power spectrum without 
small scale cutoff which is linked, for example, with the finite 
mass of DM particles, and without correction for the survival 
probability. So, it describes only the massive part of the mass 
function. 

Relations (\ref{nm1}), \&  (\ref{nm2}) characterize the mass 
distribution of dark matter clouds associated with the observed
galaxy groups and massive structure elements. They are closely 
linked with the initial power spectrum and fit reasonably well 
the observed mass distribution. For $N_{sel}\leq \langle N_{sel}
\rangle$ the incompleteness of the sample of selected clouds leads 
to rapid drops in the observed mass functions as compared with 
theoretical expectations. For filaments selected within LDRs at 
$r_{lnk}\leq 2$ the deficit of richer clouds is enhansed by the 
method of filament separation. 
However, for the largest linking lengths, $r_{lnk}=2.6 ~\&~ 3.6 
h^{-1}$Mpc, where the incompleteness and other distortions are 
minimal, the observed mass distribution is quite consistent with 
theoretical expectations. 

\section{Summary and discussion}

Statistical analysis of large galaxy redshift surveys allows us to
obtain the quantitative characteristics of the large scale galaxy
distribution, which in turn can be related to the fundamental
characteristics of the Universe and the processes of structure
formation.  The large homogeneous data set compiled in the SDSS DR1
also permits us to checking the results from analysis of the LCRS 
and the DURS and to obtain more accurate and more representative 
estimates of the main basic characteristics of the Universe.

The spatial galaxy distribution for the $S_1$ samples is
plotted in Fig. 15; galaxies in HDRs are highlighted.  

\begin{figure}
\epsfxsize=8.5cm
\hspace{1.5cm}
\epsfbox{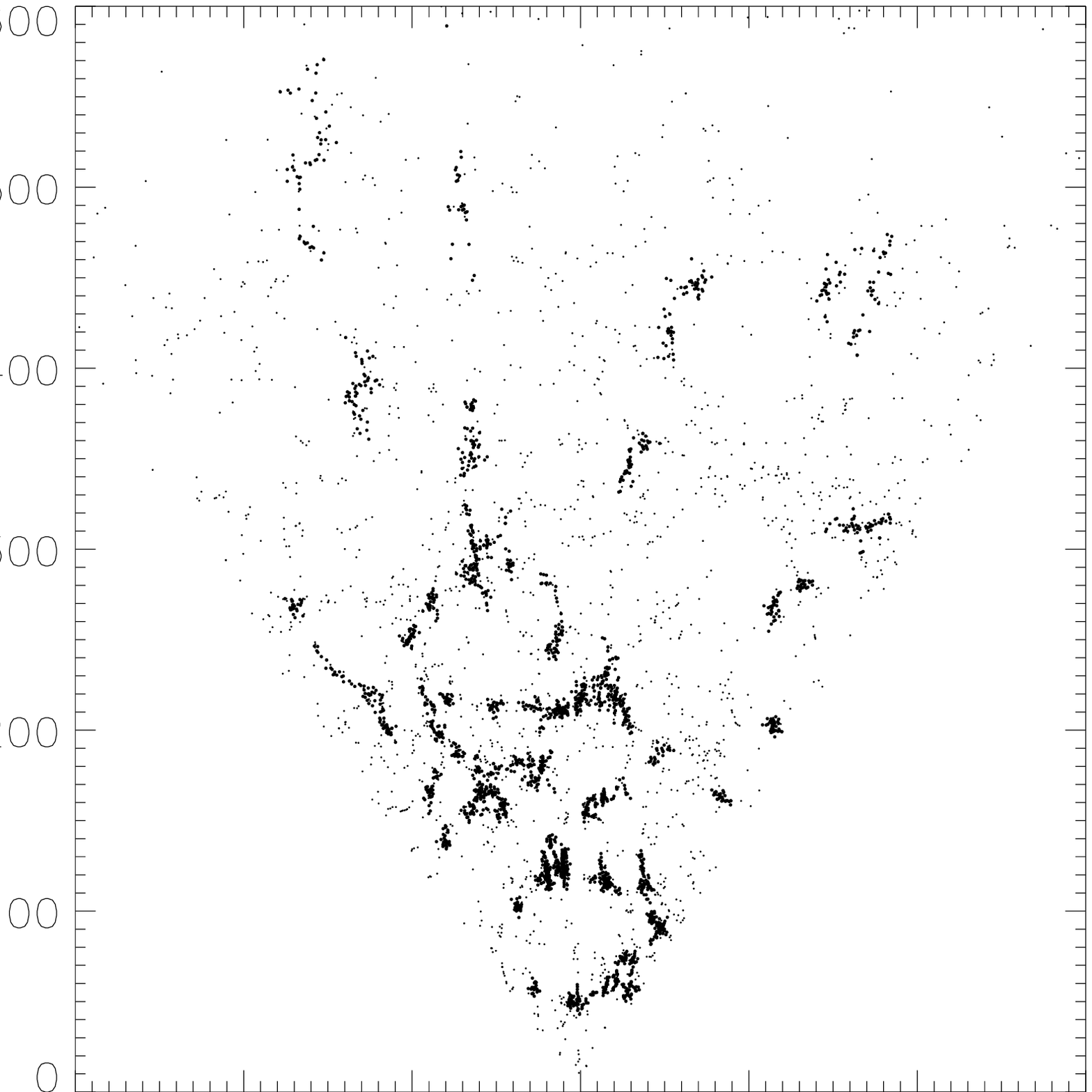}
\vspace{1.2cm}
\caption{Spatial distribution of all galaxies in the
$S_1$ sample (points) and  within the HDRs (thick points).
}
\label{S1}
\end{figure}

\subsection{Main results}

The main results of our investigation can be summarized as follows:

\begin {enumerate}
\item{}
	The analysis performed in Sec. 3 with the MST technique 
	confirms that about half of galaxies are situated within 
	rich wall--like structures and the majority of the remaining 
	galaxies are concentrated within filaments. This result 
	confirms that the filaments and walls are the main structure 
	elements in the observed galaxy distribution. Quantitative 
	characteristics of walls and filaments presented in Sections 
	3, 4 \& 5 validate this division of the LSS into these 
        two subpopulations.  
\item{}
	The main characteristics of wall--like structure elements, 
	such as the overdensity, separation distance between walls,  
	wall thickness, and the velocity dispersion within walls, 
	were measured separately for radial and, for the SDSS DR1 
	equatorial stripes, in transverse directions. 
	Comparison of these characteristics demonstrates that the 
	walls are approximately in static equilibrium, that they are 
	relaxed along their shorter axis, and that their observed 
	thickness in the radial direction is defined by the velocity 
	dispersion of galaxies. 
\item{}
	The PDF of the wall surface density is consistent with the
	simulated one and with that predicted by Zel'dovich theory 
	for Gaussian initial perturbations and CDM--like initial 
	power spectrum. The measured amplitude of perturbations 
	coincides with that expected for the spatially flat 
	$\Lambda$CDM cosmological model with $\Omega_\Lambda\approx$ 
	0.7 and $\Omega_m\approx$ 0.3\,. These results demonstrate 
	that the spatial galaxy distribution traces the dark 
	matter distribution nicely.
\item{}
	The mass distributions of groups of galaxies, filaments 
	and walls selected with various threshold overdensities
	nicely fit the joint mass function consistent with 
	the expectations of Zel'dovich theory.
	Characteristics of these groups of galaxies listed in Table 2 
	clearly demonstrate the impact of environment and interaction of 
	small and large scale perturbations. 
\item{}
	We found the typical cell size of the filamentary network to 
	be $\sim 10 h^{-1}$Mpc (Sec. 4). This estimate is consistent 
	with the one obtained previously for the LCRS \citep{DoTu:96,DoFo:01}.
\end{enumerate}

\subsection{Characteristics of the LSS}

Results obtained in Sec. 3 with the MST technique demonstrate 
that at least 80 -- 90\% of galaxies in the SDSS DR1 are 
concentrated within the filaments and walls with a variety of  
richness and overdensity forming the LSS. Comparison of the 
mean edge lengths (\ref{lmst_i}, \ref{lodr} and \ref{ludr}) 
for the full sample, HDRs and LDRs with the mean density of the 
sample (\ref{smp}) shows that the LSS occupies only roughly half 
of the volume. It also shows that the mean overdensity of the walls is larger 
than 10. The same results show that the filaments are formed 
by a system of high density groups of galaxies which contain 
$\approx$ 30\% of all galaxies. For only $\approx$ 20\% of 
galaxies do we see the actual one dimensional PDF of the MST edge 
lengths (Fig. \ref{mst}).  

There is continuous distribution in morphology and richness 
of the LSS elements. It can be roughly described as a system 
of walls randomly distributed in space with mean separations of
$\approx (50\pm 10) h^{-1}$Mpc and a filamentary network 
connected to the walls. The typical size of lower density regions,
or voids, situated between filaments and walls is estimated 
in Sec. 4 as approximately 10--12 $h^{-1}$Mpc. 

These results agree with those obtained for the mock 
catalogues simulating the galaxy distribution and indicate 
that galaxies nicely trace the spatial distribution of dominated 
dark matter. 

\subsection{Properties of walls and parameters of the initial 
power spectrum}

Walls and filaments are the largest structure elements 
observed in the Universe. In contrast to galaxies, their  
formation occurs at relatively small redshifts in course 
of the last stage of nonlinear matter clustering and is 
driven by the initial power spectrum of perturbations. 
Therefore, their properties can be successfully described 
by the nonlinear theory of gravitational instability \citep{Zeld:70} that
allows us to link them with the characteristics of the initial power spectrum. 

The interpretation of the walls as Zel'dovich pancakes has been
discussed already in \cite{ToGr:78} and in \cite{Oo:83}.
The comparison of the statistical characteristics of the
Zel'dovich pancakes for a CDM--like initial power spectrum (DD99,
DD02) with those for observed walls demonstrates that, indeed, this
interpretation is correct and, for a given cosmological model, it
allows us to obtain independent estimates of the fundamental
characteristics of the initial power spectrum.

The estimates of the mean wall surface density, $\langle q_w
\rangle$, and the amplitude of initial perturbations, $\langle
\tau_m\rangle$, listed in Table 1 are consistent with each other 
and with those found for the LCRS and DURS. They are also close 
to those found for the simulated DM distribution and for the mock 
galaxy catalogs \citep{CoHa:98} prepared for a spatially flat 
$\Lambda$CDM cosmological model with $\Omega_\Lambda=0.7$, 
$\Omega_m=0.3$ and $\sigma_8=1.05$. As was shown in Sec. 5.3 
(\ref{esim}), (\ref{pobs1}) and (\ref{pobs2}), the PDFs of both 
observed and simulated wall surface density plotted in Figs. 
\ref{qsim} and \ref{qobs} coincide with those theoretically 
expected (\ref{wq}) for Gaussian initial perturbations with a 
CDM--like power spectrum (DD99; DD02). 

Averaging of both $\langle q_w\rangle$ and $\langle\tau_m\rangle$ 
listed in Table 1 allows us to estimate the mean values as follows:
\be
\langle q_w\rangle = (0.38\pm 0.06)(\Gamma/0.2)\,,
\label{qmm}
\ee
\be
\tau_m = (0.24\pm 0.02)\sqrt{\Gamma/0.2}\,.
\label{staum}
\ee 
These values are consistent with the best estimates of the same 
amplitude \citep{Sper:03} for the $\Lambda$CDM cosmological 
model with $\Gamma=0.2$ 
\be
\sigma_8\approx 0.9\pm 0.1,\quad \tau\approx 0.22\pm 0.02\,,
\label{ampl}
\ee 

These results verify that galaxies nicely trace the LSS formed 
mainly by the dark matter distribution and the observed walls are  
recently formed Zel'dovich pancakes. They also verify the Gaussian 
distribution of initial perturbations and coincide with the Harrison 
-- Zel'dovich primordial power spectrum. 

Comparison of other wall characteristics measured in radial 
and transverse directions indicate that the walls are 
gravitationally confined and relaxed along the shorter axis. The 
same comparison allows us to find the true wall overdensity, 
wall thickness, and the radial velocity dispersion of galaxies within 
walls. As is seen from relation (\ref{phi}), these values are 
quite self--consistent. 

\subsection{Mass function of structure elements}

The samples of walls, filaments, and groups of galaxies in the SDSS
DR1 selected using different threshold overdensities allow us to
measure their mass functions, to trace their dependence on the
threshold overdensity and environment, and to compare them with 
the expectations of Zel'dovich theory.

This comparison verifies that for lower threshold overdensities for
both filaments and wall--like structure elements, the shape of the
observed mass functions is consistent with the expectations of
Zel'dovich theory. However, for high density groups of galaxies
some deficit of low mass groups caused, in particular, by selection 
effects and enhanced by the restrictions inherent in our procedure 
for group-finding leads to a stronger difference between the observed 
and expected mass functions for $M_{sel} \leq\langle M_{sel}\rangle$. 
The same factors distort the observed mass functions for groups 
selected within LDRs.  

Let us note that mass functions (\ref{nm1},\,\ref{nm2},\,\&\,
\ref{ps}) are closely linked with the initial power spectrum. This  
is manifested as a suppression of the PDFs at $M_{sel} \leq\langle 
M_{sel}\rangle$ and is proportional to
$\exp[-(M/\langle M_{sel}\rangle)^{1/3}]$
at $M_{sel} \geq\langle M_{sel}\rangle$. They differ from the mass 
function of galaxy clusters and the probable mass function of 
observed galaxies which are formed on account of multi--step merging 
of less massive clouds and are described by a power law with a 
negative power index at  $M_{sel} \leq\langle M_{sel}\rangle$ and 
an exponential cutoff $\propto \exp[-(M/\langle M_{sel}\rangle)$ 
at $M_{sel}\geq\langle M_{sel}\rangle$ \citep[see e.g.][]{SilkWh:78}.

\subsection{Interaction of large and small scale perturbations}

The data listed in Table 2 shows that the majority of high density 
groups of galaxies and the main fraction of galaxies related to 
these groups (up to 80 -- 90\%) are situated within HDRs. These 
results illustrate the influence of environment on galaxy 
formation and the clustering of luminous matter. It also indicates
the importance of interactions of small and large scale 
perturbations for the formation of the observed LSS.  
This problem was also discussed in \cite{Ee:03,Ein:03}. 

These differences are partly enhanced by the influence of 
selection effects. Indeed, the majority of HDRs are situated 
at moderate distances $D\sim 150 - 300 h^{-1}$Mpc where this 
influence is not so strong. On the other hand the LDRs also include all 
galaxies situated in the farthest low density regions. However, 
as was shown in DD02 significant differences between the
characteristics of clouds separated within HDRs and LDRs is 
seen even for simulated clustering of dark matter. 

These results are consistent with the high concentration of 
observed galaxies within the filaments and walls forming the 
LSS noted in Sec. 8.2. Bearing in mind that the mean density 
of luminous matter does not exceed 10\% of the full matter 
density of the Universe we can conclude that galaxies are 
formed at high redshifts presumably within compressed regions 
which are now seen as elements of LSS.

A natural explanation for both of these differences and the high 
concentration of galaxies within the LSS elements is the interaction 
of small and large scale perturbations when large scale 
compression of matter accelerates the formation of small 
scale high density clouds. Such interactions may also explain
the existence of large voids similar to the B\"ootes void where 
formation of galaxies has been strongly suppressed. 

\subsection{Possible rich clusters of galaxies}

The MST technique generalizes the standard `friends--of--friends' 
method of the selection of denser clouds and of probable clusters 
of galaxies. However, first attempts of such selection presented 
in Sec. 6 show that there is an essential difference between the  
selected high density clouds and x-ray sources. The nature of this 
difference is not yet clear and perhaps it will be eliminated
after the introduction of stronger criteria for the selection of 
probable clusters of galaxies from the survey. 
 
\subsection{Final comments}

The SDSS \citep{York:00,StLu:02,Ab:03}
and 2dF \citep{CoDaMa:01} galaxy redshift surveys provide 
deep and broad vistas with which cosmologists may study the galaxy 
distribution on extremely large scales --scales on which the imprint 
from the primordial fluctuation spectrum has not been erased.

In this paper, we have used the SDSS DR1 to investigate the galaxy 
distribution at such large scales.  We have confirmed our earlier 
results, based on the LCRS and DURS samples, that galaxies are 
distributed in roughly equal numbers between two different 
environments: filaments, which dominate low density regions, and 
walls, which dominate high density regions. Although different 
in character, these two environments together form a fragmented joint 
random network of galaxies -- the cosmic web.

Comparison with theory strongly supports the idea that the properties
of the observed walls are consistent with those for Zel'dovich
pancakes formed from a Gaussian spectrum of initial perturbations 
for a flat $\Lambda$CDM Universe ($\Omega_{\Lambda} \approx 0.7$, 
$\Omega_m\approx 0.3$). These results are consistent with the
estimate of $\Gamma=0.20\pm 0.03$ obtained in \cite{Pe:01}
for the 2dF Galaxy Redshift Survey \citep[see also][]{Sper:03}. 

Such analysis one allows to obtain some important basic conclusions 
regarding the properties and the process of formation and evolution 
of the large scale structure of the Universe.  With future public 
releases of the SDSS data set, we hope to refine these conclusions.

\section*{Acknowledgments}
We thank Shiyin Shen of the Max-Planck-Institut f\"ur Astrophysik 
and J\"org Retzlaff of the Max-Planck-Institut f\"ur Extraterrestrial 
Physics for useful discussions regarding this work. 

Funding for the creation and distribution of the SDSS Archive has been
provided by the Alfred P. Sloan Foundation, the Participating Institutions,
the National Aeronautics and Space Administration, the National Science
Foundation, the US Department of Energy, the Japanese Monbukagakusho,
and the Max Planck Society. The SDSS Web site is http://www.sdss.org/.
\\
The Participating Institutions are the University of Chicago, Fermilab, the
Institute for Advanced Study, the Japan Participation Group, the Johns
Hopkins University, the Max Planck Institute for Astronomy (MPIA), the
Max Planck Institute for Astrophysics (MPA), New Mexico State
University, Princeton University, the United States Naval Observatory, and
the University of Washington.
\\
This paper was supported in part by Denmark's Grundforskningsfond through 
its support for an establishment of the Theoretical Astrophysics Center.

\end{document}